\documentclass[aps,showpacs,floatfix,superscriptaddress,twocolumn]{revtex4-1}
\usepackage{amsmath,amssymb}
\usepackage{verbatim,graphicx}

\usepackage{color}
\usepackage[all]{xy}

\newcommand{\vev}[1]{\langle #1\rangle}
\newcommand{\ket}[1]{| #1\rangle}

\newcommand{\BZ}{\mathbb{Z}}

\newcommand{\beq}{\begin{equation}}
\newcommand{\beqs}{\begin{equation*}}
\newcommand{\eeq}{\end{equation}}
\newcommand{\eeqs}{\end{equation*}}

\allowdisplaybreaks

\begin{document}

\title{Universal quantum constraints on the butterfly effect}

\author{David Berenstein}
\affiliation {Department of Physics, University of California at Santa Barbara, CA 93106
and Department of Applied Math and Theoretical Physics, Wilbeforce Road, Cambridge, CB3 0WA, UK}
\author{Antonio M. Garc\'\i a-Garc\'\i a}
\affiliation{TCM Group, Cavendish Laboratory, University of Cambridge, JJ Thomson Avenue, Cambridge, CB3 0HE, UK}

\begin{abstract} 
Lyapunov exponents, a purely classical quantity, play an important role in the evolution of quantum chaotic systems in the semiclassical limit. We conjecture the existence of an upper bound on the Lyapunov exponents that contribute to the quantum motion, namely, even in the semiclassical limit
only a limited range of Lyapunov exponents, bounded from above, are important for the quantum evolution.
This is a universal feature in any quantum system or quantum field theory, including those with a gravity dual. 
It has its origin in the finite size of the Hilbert space that is available to an initial quasi-classical configuration. An upper bound also exists in the limit of an infinite Hilbert space provided that the system is in contact with an environment, for instance a thermal bath. An important consequence of this result is a universal quantum bound on the maximum growth rate of the entanglement entropy at zero and finite temperature. 
\end{abstract}


\maketitle
A central feature of classical chaos is the  {\it butterfly effect}, namely, the extreme sensitivity of the dynamics to small changes in the initial conditions. It is governed by Lyapunov exponents that control the exponential growth of the separation between nearby  classical trajectories, 
$|\delta x(t)| \simeq |\delta x(0)| \exp( \kappa t) $
where $\delta x(0)$ stands for a small change of initial conditions, 
and $\kappa$ is the largest Lyapunov exponent available to the system. 

In the semiclassical limit ($\hbar \to 0$) the classical butterfly effect also plays an important role in the quantum dynamics \cite{berman1978} for sufficiently short times $t \lesssim  t_{\rm E}$, where $t_{\rm E} \sim \log(\hbar^{-1})/\kappa$ is the Ehrenfest time defined as the time for which the initially small quantum corrections become comparable to the classical contributions.
In this region, the leading $\hbar$ correction of certain dynamical observables grows exponentially with the separation of classical trajectories $\exp( \kappa t) $ \cite{larkin1969}.
 


A paradigmatic example of this class is the rate of growth of the entanglement entropy (EE).
 Paz and Zurek \cite{zurek1994} conjectured that the slope of the linear growth of the EE in an initially coherent quantum chaotic system after coupling it to thermal bath is just the classical Kolmogorov-Sinai (KS) entropy which depends only on the Lyapunov exponents. Numerical simulations \cite{miller1998} have largely confirmed the conjecture at least in the limit of high temperatures or sufficiently small Lyapunov exponents. 
Similar results are observed even if there is no notion of temperature until after the system has scrambled from an initial semi-classical state. The EE growth rate, obtained by tracing over half of the classical  degrees of freedom of a chaotic dynamical system \cite{asplund2015} (see also \cite{latora1999}), 
is again the KS entropy and is essentially insensitive to the choice of degrees of freedom that are traced over.

Qualitatively  this can be understood by imagining two identical quantum chaotic systems $A,B$ (each with one canonical pair of degrees of freedom $x,p$) that at $t=0$ are put in contact. The entanglement entropy $S= - Tr \rho_A \log \rho_A$ is computed by integrating over the system $B$ where $\rho_A =Tr_B  \rho_{AB}$  is the reduced density for system $A$ and $\rho_{AB}$ is the total density matrix. Again assuming that the initial state for $A$ is a coherent state it is possible to prove that up to times of the order of the Ehrenfest time the entropy will grow like the logarithm of the area in units of $\hbar$. That is,  $S \sim \log (\Delta x(t) \Delta p(t)/\hbar)$.
This leading contribution to $S$, which is correct for times shorter but of the order of the Ehrenfest time,  will come from the semiclassical expressions for the uncertainty $\Delta x$ and $\Delta p$. According to the discussion above, these uncertainties will grow exponentially in time with a rate controlled by the (two) largest Lyapunov exponents of the total system $A+B$.  To a good approximation the entropy growth rate is $ \frac{\Delta S}{\Delta n} \simeq 2 \kappa_+$ where $\kappa_+$ is the largest Lyapunov exponent associated to $A$ and $B$ and $n$ stands for the number of time steps.
This result indicates that the loss of information by entanglement, a purely quantum  phenomenon, is controlled by the purely classical butterfly effect. This intuition generalizes easily to systems with many degrees of freedom, where we expect the EE to be proportional to to the volume and still to grow as a sum of the largest Lyapunov exponents \cite{asplund2015}.

Recent results in very different problems ranging from quantum information \cite{bravyi2007,acoleyen2013} and quantum chaos \cite{fujisaki2003,miller1999,alicki2004} to field theories with a gravity dual \cite{maldacena2015,hod2007} have challenged the universal validity of this semiclassical analysis. It seems that for sufficiently large $\kappa$ or low temperatures the EE growth rate saturates to a value dependent on $\hbar$ rather than to the purely classical Lyapunov exponents.

Our goal in this paper is to clarify the possible origin of the physics responsible for the breakdown of the semiclassical picture, and to study its physical consequences. We will argue that there exists a universal quantum bound, proportional to the logarithmic of the number of states, for the Lyapunov exponents that enter in the semiclassical analysis. 
This bound is non-perturbative in $\hbar$ and does not need a temperature to manifest itself. The  origin of the bound lies in the finiteness of the Hilbert space that is available to a dynamical system with a fixed semi-classical initial condition.
An important consequence of this result, that we discuss in detail, is that the growth rate of information loss, as measured by the entanglement entropy, has a universal upper quantum bound. 
The effect of the environment can also break down the semiclassical prediction for this growth rate even if the full Hilbert space is not finite. For the case of thermal environments we propose, based on black hole physics and causality constraints, a universal bound for the EE growth rate.

Next we give general arguments for the existence of bounds in the Lyapunov exponents and the growth rate of the entanglement entropy.\\
 {\it Uncertainty and universal bounds on the butterfly effect.-}\\
A natural set of observables to investigate these universal bounds are the
commutators that appear in the generalized uncertainty relation. Consider applying this uncertainty relation to the commutator of an operator at two points of  time,  
\begin{equation}
\Delta x_n \Delta x_0 \geq |[\hat x_n, \hat x_0]|/2 \label{eq:unc}
\end{equation}
where for convenience we have switched to discrete time labelled by $t=0,n$.
Assuming that the initial state at $t=0$ is coherent $\Delta x_0 \approx \sqrt{\hbar}$ and computing the commutator to leading order in $\hbar$, 
\begin{equation}
\Delta x_n \geq |[\hat x_n, \hat x_0]|/(2\Delta x_0) \approx \hbar^{-1/2} \vev{  \{ x_n,x_0\}_{P.B.} },
\end{equation}
we find that it depends on the Poisson Bracket between functions at different times.
 The right hand side is proportional to the Jacobian $\partial x(t)/\partial x(0)$ 
 which for chaotic systems leads to an exponential growth of the Poisson bracket \cite{larkin1969,berman1978} for times less than the Ehrenfest time, 
\begin{equation}
\Delta x_n \geq |[\hat x_n, \hat x_0]|/{2\Delta x_0} \approx \vev{  \{ x_n,x_0\}_{P.B.} } \approx \sqrt{\hbar} e^{\kappa_+ n},\label{eq:lyap}
\end{equation}
where $\kappa_+$ is the largest positive Lyapunov exponent. 

The growth of the quantum uncertainty seems to be solely controlled by the classically chaotic dynamics. 
We interpret this to signify that the butterfly effect in quantum mechanics is the (exponential) deterioration in the capacity to do detailed 
predictions for quantum observables as time increases.  Let us consider the growth of uncertainly after a single step. The  uncertainty in $x_1$ has to be smaller than the maximum possible uncertainty in the system $\Delta x_1 < \Delta x_{max}\simeq A$, which cannot be larger than the typical system size $A$. This size does not depend on $\hbar$ in a semiclassical treatment therefore,
\begin{equation}
A \sqrt{\hbar} >\Delta x_0 \Delta x_1 \geq |[\hat x_1, \hat x_0]|/2 \approx \hbar e^{\kappa_+}, \label{eq:ineqsandwich}
\end{equation}
where we are assuming that the initial state is coherent  $\Delta x_0 \sim \sqrt{\hbar}$. Here we are only counting the powers of $\hbar$ as $\hbar \to 0$. To get the units to work right, a prescription for how to precisely choose the coherent states needs to be implemented (this is less of an issue if we use a canonical pair of variables $x,p$, as the volume occupied by them satisfies exactly that 
$\Delta x_0\Delta p_0= \hbar/2$).

For both inequalities to be valid in \eqref{eq:ineqsandwich}, the Lyapunov exponent must be bounded $\kappa_+ < B \log (\hbar^{-1})$ with a prefactor $B$ of order one that does not depend on $\hbar$. 


This bound in the quantum butterfly effect is also closely related to the finite dimension $N \sim \Delta x_{max} \Delta p_{max} /\hbar$ of the Hilbert space for a single degree of freedom in quantum mechanics (basically, the volume of phase space of the degree of freedom  in units of $\hbar$). This property leads to a finite time to fill the available phase space.  As a consequence the bound on the Lyapunov exponent can be cast as, 
$ 2 \kappa_+ < \log N \approx \log (\int d x d p /\hbar)$. Here, the more precise version is that the sum of the two largest Lyapunov exponents is bounded by $\log(N)$, but if they're similar, we get the bound. Otherwise, the bound deteriorates as we add more degrees of freedom.
That the bound is related to finite $N$ is also clear from the calculation of the entanglement entropy for splitting a system from many degrees of freedom into two subsystems $A,B$. The entropy production per unit time must be less than the maximal entropy of a density matrix on the first system $B$. The maximum of the entanglement entropy for such a splitting is limited by $\log N$ so 
\begin{equation}
\kappa_+ < {\log N \over 2d} \label{eq:b2}
\end{equation}
which is the same bound as the one obtained above, where in this case $d$ is the spatial dimensionality, with another $d$ arising from the conjugate momenta.
Based on the above discussion we propose that these bounds are universal for any quantum system with a chaotic classical counterpart and a finite Hilbert space at zero or at finite temperature. By universal we mean that it is always proportional to the logarithm of the Hilbert space dimension, a dimensionless quantity. The prefactor, that sets the time scale of the Lyapunov exponent, is in general non universal. Generically, one can argue that it is determined by other time scales in the problem, like the time it takes to explore the range of a typical single variable, or the typical time between collisions in a gas.
We also stress that the bound does not signal the complete breakdown of the semiclassical approximation, as the quantum correlation functions of interest still grow exponentially, but rather limits the time scale for which the semi-classical formalism is valid.


An immediate consequence of this result is a universal bound on the growth rate of the EE. As was mentioned previously, in the semiclassical limit, the growth of the EE is linear in time and proportional to the sum of the Lyapunov exponents. The growth rate is limited by the maximum Lyapunov exponent consistent with Eq.(\ref{eq:b2}). Quantum mechanical effects induce entanglement and decoherence but, importantly, also limit its growth. We show next how the bound arises in a specific example.\\ 
{ \it  Quantum cat maps and growth of EE in a toy model field theory.-}\\
A  classically chaotic map is quantized \cite{berman1978} by writing down the evolution matrix in discrete time acting on an initially coherent state. Quantum operators are then expressed as c-numbers describing the initial coherent state. We want a system where the Lyapunov exponents are calculable. The simplest such model is an iterated linear transformation acting on $x,p$. However, we also need the phase space to be compact. This forces us to consider a system where the $x,p$ variables  are periodic.

An example is the algebra of the so called fuzzy torus generated by two unitary operators $U,V$ subject to $U^N=V^N=1$ and $UV= \exp(2\pi i/N) VU$.
Here $U\simeq \exp(2 \pi i\hat x)$, and $V\simeq \exp(2\pi i \hat p)$, with $\hat x,\hat p$  position and momentum operators of period one and commutator $\sim \hbar \simeq (2\pi N)^{-1}$. 
The unitary irreducible representations of the algebra act therefore on a Hilbert space of dimension $N$.
If $x,p$ are c-numbers associated to variables on a torus, the mapping of the torus can be realized by 
\begin{equation}
\begin{pmatrix} x\\
p
\end{pmatrix}
\to\begin{pmatrix} a& b\\
c&d \end{pmatrix}  \begin{pmatrix} x\\
p\end{pmatrix}= M  \begin{pmatrix} x\\
p\end{pmatrix}
\end{equation}
with $ab-cd=1$. For $a=2, b=c=d=1$ this is the Arnold cat map, an exactly solvable chaotic system. 
The Lyapunov exponents
are computed by taking the logarithm of the eigenvalues of $M$. These are given by $\lambda_{1,2} =\kappa_{\pm}=\pm \kappa$. In order to quantise it, we associate the same matrix to an action that acts as an automorphism of the algebra generated by $U,V$, where 
\begin{equation}
U\to \eta U^a V^b \quad V\to \eta' U^c V^d
\end{equation}
and $\eta, \eta'$ are phases that are determined by  $( \eta U^a V^b)^N= \eta^N U^{Na} V^{Nb} \exp(\pi i a b (N-1))=1 $, so $\eta, \eta'$ only need appear if $N$ is even.
Let us call the unitary  operator that implements that automorphism $\Gamma$.
Because $U^N=1$, the $a,b, c,d$ only need to be defined modulo $N$. Thus, at finite $N$ we really only have an action by an element of $SL(2,\BZ_N)$, rather than $SL(2,\BZ)$, but there is a standard map from $SL(2,\BZ)\to SL(2,\BZ_N)$.

Now, let us return to our example of the fuzzy torus. 
We can now study commutators for $U$ displaced in time.
A straightforward computation shows that under iteration,
\begin{equation}
U_n U -U U_n =\left (1- \exp\left[\frac{2\pi i}{N} (M^n)_{12}\right]\right)  U_n U\label{eq:unc2}
\end{equation}
Now, we  insert this  expression for the commutator on the right hand side  of equation \eqref{eq:unc}. To make the operators hermitian we can take $U+U^{-1}$ for example, but the estimate is essentially the same. We see that in this case the role of $\hbar \exp(\kappa_+n)$ is played by $ (1- \exp(\frac{2\pi i}{N} (M^n)_{12}))\simeq -\frac{2\pi i}{N} (M^n)_{12} $, which at large $N$ tell us that $\hbar \simeq (2\pi N)^{-1}$ and that the Lyapunov exponents are the eigenvalues of $M$. 
We observe that if $N$ is large, the right hand side of \eqref{eq:unc2} grows exponentially for small enough values of $n$: when $ 2\pi (M^n)_{12}/N <1$. We want this inequality to be true at least for $n=1$. The inequality is violated if $M$ has large eigenvalues, so we argue that quantum mechanics puts a limit on scrambling.

More precisely, we can argue that a minimal uncertainty package in $U,V$ centered around phases $(\theta, \phi)$ in the torus should be such that the uncertainty after the next iteration of the dynamics is $\Delta U_1<1$. Such a state $\ket\psi$ is defined by minimizing the hermitian operator $(U-\exp(i\theta))^\dagger(U-\exp(i\theta))+(V-\exp(i\phi))^\dagger(V-\exp(i\phi))$ (see for example \cite{Ishiki:2015saa}). The state $\ket\psi$ has an uncertainty in $U,V$ or roughly $1/\sqrt N$ in each direction. The left hand side of equation \eqref{eq:unc2}
is also given by
$[U_n- exp(i\theta_n),U-\exp(i\theta) ]=U_n U -U U_n$
where $\theta_n\simeq M^n( \theta,\phi)^T$. Then the left hand side can be related to the uncertainties in a straightforward way.
This gives us a bound when we plug it in
$\Delta U_1 \frac 1{\sqrt N} \geq \frac 1  N \exp(\lambda_+)$ 
or equivalently, $\lambda_+\leq \log(\sqrt N)$, the same bound as \eqref{eq:b2}.

We now extend this further to systems that have multiple coordinates, but that are still similarly soluble, namely, a toy model of an interacting quantum field theory on a 1-D lattice.  If we have a lattice system with a Hamiltonian, the Lieb Robinson bound \cite{lieb1972} suggest that we can pretend that time is discrete with a time step that is the effective light-crossing time per site. This is enough to motivate our model: take a  lattice with a set of $U,V$ matrices at each site.
Consider $U_{(a)}\to U_{(a)} V_{(b)}$, $U_{b}\to U_{b} V_{(a)}$ for some $a\neq b$.  A direct computation shows that it is an automorphism of the operator algebra between two sites similar to hopping between $a$ and $b$. We do this operation between every pair of neighboring sites on the (1-D) lattice and then follow this by a fixed quantum cat map at each site. This generates a system with nearest neighbor hopping and local scrambling.
This can still be thought of as a linear transformation for the log of the matrices interpreted as classical coordinates.
The analog of the matrix $M$ is now a collection of banded matrices
\begin{equation}
\tilde M_{nn}= \begin{pmatrix} 
\ddots&1&0&0&0&\\
0&1 & 0 & 0&0&\\
0&0&1 &1&0&\\
0&0&0&1&0&\\
0&1&0&0&1&\\
&&&&&\ddots
\end{pmatrix} \quad \tilde M_{\Gamma}= \begin{pmatrix}\ddots&&\\
&M& 
\\ & &
\ddots \end{pmatrix} 
\end{equation} 
and when we multiply these two $M_{tot}=\tilde M_{nn}.\tilde M_{\Gamma}$, we still get a banded matrix and it is periodic. This has important consequences for the eigenvalues: they give rise to a band structure similar to a periodic potential with nearest neighbor hopping. 

For field theories it is  very interesting to understand how much entanglement entropy is produced between two subsystems on iteration of the transformation associated to $M_{tot}$. If we have $k$ sites, one subsystem will consist of $m$ consecutive sites, and the other subsystem will be the rest of the $k-m$ sites. If $m<<k-m$, the entanglement entropy  rate is the sum of the logarithm of the $2m$ largest
eigenvalues of $M_{tot}$ \cite{asplund2015} (each site counts as two degrees of freedom: one $p$ and one $x$), which are all roughly the same and close to the maximum eigenvalue of $M_{tot}$, $\lambda_{max}= \exp(\kappa_{max})$. This is a consequence of the band structure of the eigenvalues of $M_{tot}$. This gives us
\begin{equation}\label{eet}
\frac{\Delta S}{\Delta n} \simeq 2 m \kappa_{max}
\end{equation}
and now we need this entropy per unit time to be less than the maximal entropy for a density matrix on the first system, otherwise, the right hand side overshoots a lot. This is, we get that 
\begin{equation}
m \log(N) < 2 m  \kappa_{max}
\end{equation}
which is again, the same bound from equation \eqref{eq:b2}. Notice that this means that scrambling is proportional to the volume $m$. 
So (entanglement) scrambling can be considered to be taking place locally.\\ 
{\it Discussion.-}\\
We now show that the proposed bound on the growth rate of the EE is fully consistent with recent results in holography, quantum information and quantum chaos.
A similar growth like volume was found  \cite{Balasubramanian:2010ce} holographically though for times much shorter than the Ehrenfest time. The recently proposed {\it entanglement tsunami} idea \cite{Liu:2013iza,Liu:2013qca,Casini:2015zua,avery2014} predicts bounds on entanglement rates proportional to the area bounding the regions. There is no overlap with our results as the local saturation of the entropy in this case is expected for shorter times smaller than the light crossing time of a sufficiently extended region. At local saturation the semiclassical approximation, a key ingredient in our approach, breaks down because uncertainties grow too large. The rest of the evolution occurs in the quantum regime. The scrambling rate in this case is bounded by a different reason: the ability of a single site to act as a channel with fixed capacity to connect its neighbors.\\
Bounds on the growth rate of the EE that depend on the logarithm of the dimension of the Hilbert space have also been reported in the quantum information literature though without an explicit relation to classical Lyapunov exponents and general uncertainly relations. It was conjectured \cite{bravyi2007}, and recently proved \cite{acoleyen2013}, that the growth of EE is bounded by the log of the size of the Hilbert space of the part of the system that acts as a reservoir. 
A bound in the growth rate of the EE has also been observed in quantum system with a strongly chaotic classical counterpart, such as a kicked top, put in contact with a finite dimensional environment modelled by another chaotic system \cite{fujisaki2003}. In this case the bound is not saturated as its dependence  with the Hilbert space dimension is faster than logarithmic. 
Similarly, in the context of chaotic holographic matrix models, that in the thermodynamic limit describe quantum gravity, an upper bound  \cite{HanadaKITP} to the equivalent of $\kappa_{max}$ has been recently found numerically though its dependence on the matrix size has not yet been established. 
We note that the maximum growth rate $2 m \kappa_{max}$ in this toy model of quantum gravity should be interpreted as the universal scrambling of $m$ different degrees of freedom near an extremal black hole horizon: each degree of freedom has the same scrambling controlled by $\kappa_{max}$. This meshes well with the idea that the blueshift due to gravity has the same strength for all matter.\\  
{\it Bound of the EE growth rate in thermal systems.-}\\
So far our analysis is restricted to systems with a finite Hilbert space. However on physical ground we expect that bounds must also exist for continuous systems with an arbitrarily large number of states available. For instance we expect that the Lyapunov exponent, that has units of inverse time, must be bound by the uncertainty relation, 
$
\kappa_+ < \Delta E /\hbar
$
where $\Delta E$ is a non-universal energy scale that depends on the system in question. 
For larger $\kappa_+ \geq \kappa_{\rm max} = \Delta E/\hbar$ the classical Lyapunov exponent does not control the quantum dynamics even in the semiclassical limit though the predicted exponential growth of certain correlation functions, of semiclassical origin, is still valid up to the Ehrenfest time. This is fully consistent with the recent conjecture  \cite{maldacena2015} that, for systems coupled to a thermal bath the parameter $\lambda_{\rm Q}$, that controls the exponential grow of certain commutators around the Ehrenfest time, has an universal upper bound $
\lambda_{\rm Q}\leq \frac{2\pi k_B T}{\hbar}$ in the semiclassical limit. This is the expected behaviour in thermal field theories with a gravity dual where membrane paradigm ideas \cite{susskind1995,barbon2011,thorne1986} suggest an exponential growth of the uncertainty, controlled by the blue-shift and bound by causality, of an infalling observer \cite{Shenker:2013pqa} approaching the horizon. 
Next we employ similar ideas to propose a universal bound on the EE growth rate also exists in systems coupled to a thermal bath. 

We study the growth of EE of a test particle that it is put in contact with a strongly coupled plasma, with a large $N$ gauge group, at temperature $T$. The gravity dual of this system is a Nambu-Goto string going from the horizon to the boundary. However, see \cite{deboer2009} for a detailed derivation, the low energy effective gravity dual, valid for most purposes, is just a test particle close to the so called stretched horizon \cite{thorne1986,susskind1995}. The particle-plasma scattering corresponds in the gravity dual with a scattering event in the stretched horizon followed by gliding in the thermal atmosphere of the blackhole \cite{barbon2011}, described by Rindler geometry, interrupted sometimes by reflection in the AdS boundary. With these simple ingredients we estimate the growth of the EE.

Before the scattering starts the probe is in a coherent state  $\Delta x(0) \Delta p(0) \approx \hbar$. The associated reduced density matrix is Gaussian. The rate of loss of information will have a maximum for time intervals for which the uncertainty in momentum or position grows very fast. The saturation of the EE will occur when the initially localized particle is fully spread in the stretched horizon \cite{susskind1995}. For a static observer the particle is accelerated as it falls toward the horizon. For a Schwarzschild geometry the momentum increases as 
$p \sim e^{t/4MG}$ where $G$ is the gravity coupling constant and $M$ the mass of the black hole. For short times, the spread in space $\Delta x^2 \propto \log p \propto t$. This is valid for densities much smaller than the Planck density. However, as the particle approaches the stretching horizon, the Rindler geometry applies and $\Delta x^2 \propto p \sim G {\rm e}^{t/4MG}$ where $G \propto 1/N^2$ is the gravity coupling strength.
The entanglement entropy before saturation is simply, 
\begin{equation}\label{st}
S \sim \log (\Delta x(t) \Delta p(t)) \sim t/4MG = t/\tau
\end{equation}
where $\tau = \hbar/2\pi k_B T$.
Interestingly \cite{susskind1995} a growth faster than $e^{t/4MG}$ violates causality. 
This encourages us to conjecture that $1/\tau$ is the maximum growth rate of the EE in any system in contact with a thermal bath. 
 


This bound is consistent with previous results in fermionic thermal systems \cite{song2012} where the growth rate  $\sim \hbar /k_B T$ depends on the interaction but it is always below the bound (\ref{st}). Another evidence comes from the breaking of the Zurek-Paz conjecture \cite{zurek1994} for sufficiently low temperatures where the growth rate saturates \cite{miller1999} to a value dependent on the thermal environment. 
Similarly it has been conclusively shown \cite{alicki1996,alicki2004,alicki2007,pogorzelska2007} that, for sufficiently large Lyapunov exponents, the growth rate of the EE for certain quantum maps is controlled by a generalized KS entropy of the environment \cite{pogorzelska2007}, in this case non-thermal, and not by the butterfly effect.



 
In summary we conjecture that there exists an universal bound on the Lyapunov exponents, and in general on classical information,  that enter in semiclassical expressions of quantum observables.
The bounds are related to uncertainty constraints and to the effect of the environment in the semiclassical evolution.  A notable consequence of the conjecture is a universal upper bound on the growth rate of the EE for systems with a finite Hilbert space or in contact with a thermal bath. Quantum mechanical effects induce entanglement but interestingly also limit its growth.  

\acknowledgements{D. B. would like to thank M. Hanada, S. Shenker  and M. Srednicki for useful comments. A. M. G. acknowledges partial support from
EPSRC, grant No. EP/I004637/1 and thanks the Galileo Galilei Institute for Theoretical Physics for the hospitality and the INFN for partial support during the completion of this work.
D.B. work  supported in part by the department of Energy under grant {DE-SC} 0011702. The research leading to these results has received funding from the European Research Council under the European Community's Seventh Framework Programme (FP7/2007-2013) / ERC grant agreement no. [247252].
}
\bibliography{library1}

\end{document}